\documentclass[a4paper]{article}
\usepackage{amsmath}
\usepackage{amssymb}
\usepackage{graphicx}
\usepackage{picins}
\usepackage{fullpage}
\usepackage{hyperref}

\newcommand{\R}{\mathbb{R}}
\newcommand{\G}{\mathcal{G}}
\newcommand{\Ck}{\mathcal{C}_{k}}
\newcommand{\CC}{\mathcal{C}}
\newcommand{\vt}{t}
\newcommand{\vs}{s}
\newcommand{\w}{w}
\newcommand{\FAM}{\mathcal{P}}
\newcommand{\err}{\varphi}
\newcommand{\errk}{\varphi_{k}}
\newcommand{\vst}{\psi}
\newcommand{\aper}{\alpha}
\newcommand{\ERR}{\sigma}
\newcommand{\D}{D_\ERR}
\newcommand{\peri}{\mathrm{peri}}

\author{Hee-Kap Ahn%
  \thanks{Dept.~of Computer Science and Engineering,
    Sejong University, Seoul, Korea.
    Email:~{heekap@gmail.com.}}
  \and 
  Sang Won Bae%
  \thanks{Division of Computer Science, Korea Advanced Institute of
    Science and Technology, Korea.
    Email:~\{swbae,otfried\}@tclab.kaist.ac.kr.}
  \and 
  Otfried Cheong%
  \footnotemark[3]
  \and 
  Joachim Gudmundsson%
  \thanks{National ICT Australia Ltd, Sydney, Australia.
    NICTA is funded through the
    Australian Government's Backing Australia's Ability initiative,
    in part through the Australian Research Council.
    Email:~joachim.gudmundsson@nicta.com.au.}}

\date{}

\title{Aperture-Angle and Hausdorff-Approximation of Convex Figures%
  \thanks{This research was supported by the Korea Research Foundation.}}

\newtheorem{theorem}{Theorem}
\newtheorem{lemma}{Lemma}

\newtheorem{corollary}{Corollary}

\newtheorem{conjecture}{Conjecture}

\makeatletter
\newbox\ProofSym
\setbox\ProofSym=\hbox{%
\unitlength=0.18ex%
\begin{picture}(10,10)
\put(0,0){\framebox(9,9){}}
\put(0,3){\framebox(6,6){}}
\end{picture}}
\newenvironment{proof}[1][Proof.]{\O@proof{#1}}{\O@endproof}
\def\O@proof#1{\trivlist
   \@topsep\z@\@topsepadd\smallskipamount%
   \@ifstar{\item[]}{\item[\hskip\labelsep\it #1 ]}}
\def\O@endproof{\hfill\copy\ProofSym\linebreak\endtrivlist}
\makeatother

\let\geq\geqslant
\let\leq\leqslant

\makeatletter
\partopsep\z@ \textfloatsep 10pt plus 1pt minus 4pt
\def\section{\@startsection {section}{1}{\z@}{-3.5ex plus -1ex minus
-.2ex}{2.3ex plus .2ex}{\large\bf}}
\def\subsection{\@startsection{subsection}{2}{\z@}{-3.25ex plus -1ex
minus -.2ex}{1.5ex plus .2ex}{\normalsize\bf}}
\def\@fnsymbol#1{\ensuremath{\ifcase#1\or *\or 1\or 2\or 3\or 4\or 5\or
    6\or 7\or 8\or 9\else\@ctrerr\fi}}
\makeatother

\begin{document}
\maketitle

\begin{abstract}
  The \emph{aperture angle} $\aper(x, Q)$ of a point $x\not\in Q$ in
  the plane with respect to a convex polygon~$Q$ is the angle of the
  smallest cone with apex~$x$ that contains~$Q$.  The \emph{aperture
  angle approximation error} of a compact convex set $C$ in the plane
  with respect to an inscribed convex polygon $Q \subset C$ is the
  minimum aperture angle of any $x \in C \setminus Q$ with respect
  to~$Q$.  We show that for any compact convex set $C$ in the plane
  and any $k > 2$, there is an inscribed convex $k$-gon $Q \subset C$
  with aperture angle approximation error $\big(1 -
  \frac{2}{k+1}\big)\pi$.  This bound is optimal, and settles a
  conjecture by Fekete from the early~1990s.

  The same proof technique can be used to prove a conjecture by Brass:
  If a polygon $P$ admits no approximation by a sub-$k$-gon (the
  convex hull of $k$ vertices of $P$) with Hausdorff distance~$\ERR$,
  but all subpolygons of $P$ (the convex hull of some vertices of $P$)
  admit such an approximation, then $P$ is a $(k+1)$-gon.  This
  implies the following result: For any $k > 2$ and any convex polygon
  $P$ of perimeter at most~$1$ there is a sub-$k$-gon $Q$ of $P$ such
  that the Hausdorff-distance of $P$ and $Q$ is at
  most~$\frac{1}{k+1}\sin\frac{\pi}{k+1}$.
\end{abstract}

\section{Introduction}

\parpic[r]{\includegraphics{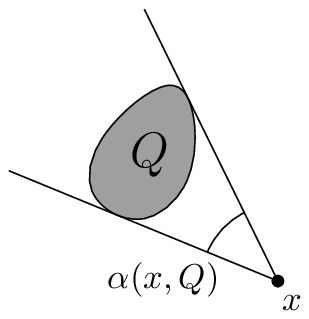}}
\noindent Let $Q$ be a compact set in the plane, and let $x$ be a
point outside~$Q$.  The \emph{aperture angle} $\aper(x, Q)$ of $x$
with respect to~$Q$ is the angle of the smallest cone with apex~$x$
that contains~$Q$ (that is, its boundary consists of two rays
emanating from $x$ tangent to~$Q$).  The aperture angle plays a role
in various applications related to sensing, and has been studied in a
number of papers.  Bose et al.~\cite{bhosg-aaop-02}, for instance,
consider two disjoint convex polygons $P$ and $Q$ in the plane, and
give algorithms to compute the maximum aperture angle and the minimum
aperture angle with respect to $Q$ when $x$ is allowed to vary in~$P$.
Hoffmann et al.~\cite{hikk-maar-98} introduce the \emph{angle hull} of
a connected region~$Q$ inside a simple polygon~$P$, consisting of all
those points~$x \in P$ with $\aper(x, Q)$ at least a given angle. They
give bounds on the length of the boundary of the angle hull, and apply
this to the problem of exploring an unknown environment.  Cheong and
van Oostrum~\cite{co-rpdu-01} give bounds on the complexity of the
angle hull of a convex polygon in a polygonal environment, and apply
this to the problem of motion planning under directional uncertainty.

We consider the problem of placing $k$ sensors or transmitters in a
compact convex room~$C$.  A point in $C$ is covered perfectly if it
lies inside the convex hull $Q \subset C$ of the sensors.  However, if
$C$ has more than $k$ vertices or even a smooth boundary, we must have
$Q \subsetneq C$, and it is not possible to achieve this for all
points of~$C$.  For points $x\in C \setminus Q$, we would like to
maximize the coverage by the sensors, and measure this using the
aperture angle~$\aper(x, Q)$.  For given $C$ and $Q \subset C$, let us
denote the worst coverage as
\[
\aper(C, Q) := \min_{x \in C} \aper(x, Q),
\]
where we set $\aper(x, Q) = \pi$ for $x \in Q$.  Since $C$ is compact
and $\aper(x, Q)$ is continuous, the minimum is indeed attained in~$C$
and this is well-defined.  We are looking for the best placement of
$k$ sensors, so we seek to maximize $\aper(C, Q)$ over all convex
$k$-gons inscribed in~$C$.  Let us denote this quantity as $\aper(C,
k)$, defined formally as
\[
\aper(C, k) := \max_{Q \in \Ck(C)} \aper(C, Q),
\]
where $\Ck(C)$ is the family of convex $k$-gons inscribed to~$C$.  In
other words, we study the approximation of convex sets by inscribed
$k$-gons with respect to the ``aperture-angle distance.''  This
distance measure is attractive as it is naturally scale-invariant,
without needing to be normalized by some global property of $C$ (such
as its perimeter or area).  We are now interested in the following
question: Given $k$, what aperture angle can we \emph{guarantee} for
any possible compact convex~$C$?  In other words, we ask for the
following quantity
\[
\aper(k) := \inf_{C \in \CC} \aper(C, k)
= \inf_{C \in \CC} \max_{Q \in \Ck(C)} \min_{x \in C} \aper(x, Q),
\]
where $\CC$ is the family of compact convex figures in the plane.
This question was first asked by Fekete in 1990, and circulated at
several open problem sessions in the early
1990s~\cite{sandor-personal}.  An upper bound for $\aper(k)$ is given
by the regular convex $(k+1)$-gon~$P_{k+1}$. Since any $k$-gon $Q$
inscribed in $P_{k+1}$ must ``miss'' a vertex of $P_{k+1}$, we have
$\aper(P_{k+1}, Q) \leq \big(1-2/(k+1)\big)\pi$ (the interior angle at
each vertex of $P_{k+1}$).  For a lower bound, we can walk around the
boundary of a given~$C$ and place a vertex of $Q$ whenever the tangent
direction has changed by~$2\pi/k$.  It is easy to see that this
achieves $\aper(C, Q) \geq (1-2/k)\pi$, and so we have $\aper(k) \geq
(1-2/k)\pi$.

A discrepancy between the two bounds remained, and Fekete conjectured
that the upper bound is correct:
\begin{conjecture}
  \label{con:aperture}
  For any $k \geq 2$, the smallest value of $\aper(k)$ is achieved by
  the regular $(k+1)$-gon, and we therefore have
  \[
  \aper(k) = \big(1-\frac{2}{k+1}\big)\pi.
  \]
\end{conjecture}
Fekete already showed that his conjecture holds for $k=2$ and $k=3$,
and experiments in Jenkner's Master thesis~\cite{j-akk-97} indicate
that it should hold for general~$k$.  The problem was also published
by Brass and Lassak~\cite{bl-pat-01}, and it appears again as
Problem~5 in Section~11.3 of Brass et al.'s encyclopedic collection of
research problems in discrete geometry~\cite{bmp-rpdg-05}, with a
short proof of the case $k \in\{2,3\}$.

If $P$ is a \emph{convex polygon}, then it is known that $\aper(P, k)$
can be attained by an inscribed \emph{subpolygon}.  Here, a
\emph{subpolygon} of $P$ is the convex hull of a subset of $P$'s
vertices.  This provides for an interesting similarity between
Conjecture~\ref{con:aperture} and the following conjecture by Brass on
Hausdorff approximation by subpolygons:
\begin{conjecture}
  \label{con:hausdorff}
  Let $\FAM$ be a family of convex polygons in $\R^{2}$ that is closed
  under taking subpolygons. If $\FAM$ has an element that is hardest
  to approximate by its $k$-vertex subpolygons with respect to the
  Hausdorff metric, then one can also find a $(k+1)$-gon in $\FAM$
  with this property.
\end{conjecture}
Conjecture~\ref{con:hausdorff} appears as Conjecture~5 in Section~11.5
of Brass et al.~\cite{bmp-rpdg-05}.  It was first suggested by Brass
in 2000~\cite{b-aps-00}. 

In this paper, we prove both Conjecture~\ref{con:aperture}
and~\ref{con:hausdorff}.  Our proof relies on a combinatorial analysis
of worst-approximable polygons in Section~\ref{sec:properties}. Here,
a polygon~$P$ is \emph{worst-approximable} if every proper subpolygon
$R$ of $P$ admits a better approximation by $k$-vertex subpolygons
than $P$ does.  Our analysis in Section~\ref{sec:properties} makes no
use of the geometry of the problem, and applies to any approximation
measure that is \emph{monotone} in the sense that ``growing'' $Q
\subset P$ cannot increase the approximation error (a formal
definition of monotonicity is in Section~\ref{sec:preliminaries}).

In Section~\ref{sec:aperture} we bring back in the geometry of the
aperture angle problem, and prove that a convex polygon that is
worst-approximable under aperture-angle approximation is in fact a
$(k+1)$-gon.  This implies a positive answer to
Conjecture~\ref{con:aperture} for the case of polygons.  The case of
arbitrary compact convex sets is then proven using a limit argument.

Similarly, we study the geometry of the Hausdorff problem in
Section~\ref{sec:hausdorff} and prove that any polygon that is
worst-approximable under Hausdorff approximation is again a
$(k+1)$-gon.

As an application of this result, we finally consider the family
$\FAM$ of convex polygons with unit perimeter.  We show that the
$(k+1)$-gon $P$ in $\FAM$ that admits the worst approximation by a
$k$-vertex subpolygon is the regular $(k+1)$-gon.  Together with our
result on Conjecture~\ref{con:hausdorff}, this implies that for every
convex polygon $P$ with unit perimeter and every $k > 2$ there is a
$k$-vertex subpolygon $Q$ of $P$ such that the Hausdorff distance
between $P$ and $Q$ is at most $\frac{1}{k+1}\sin\frac{\pi}{k+1}$.
This result is the ``subpolygon version'' of a result by Popov, who
had proven that any convex figure $C$ of perimeter one admits an
inscribed $k$-gon $Q$ with Hausdorff-distance at most
$\frac{1}{2k}\sin\frac{\pi}{k}$~\cite{p-acs-70}.  Popov's result is
not known to be tight.  Popov conjectured that the regular $(k+1)$-gon
is the worst case~\cite{p-acb-68,p-acs-70}.  Our result shows that the
equivalent statement is true for approximation by subpolygons.

\section{Preliminaries}
\label{sec:preliminaries}

\begin{figure}[htb]
  \centering
  \includegraphics[height=3cm]{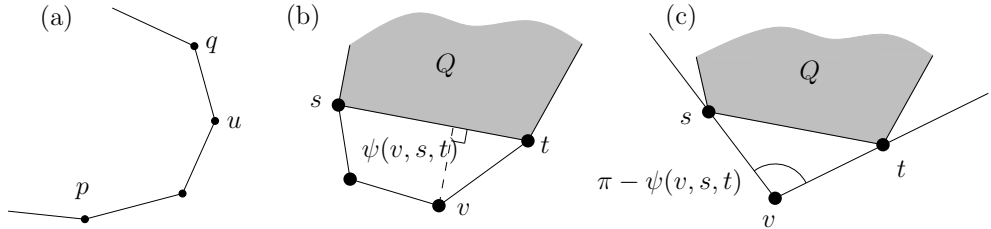}
  \caption{(a) $u$ lies between $p$ and $q$, that is,
  $p \prec u \prec q$. (b) Illustrating the Hausdorff
  distance, and (c) the aperture-angle.}
  \label{fig:preliminaries}
\end{figure}

Let $P$ be a convex $n$-gon and let $V$ be the set of its
vertices. For three vertices $p, u, q \in V$, we say that \emph{$u$
lies between $p$ and $q$} and write $p \preceq u \preceq q$ if a
counter-clockwise traversal of~$P$ starting at $p$ encounters $u$
before~$q$ (and~$u$ is allowed to coincide with $p$ or $q$).  If
we do not allow $u$ to coincide with $p$, we write $p \prec u \preceq
q$, see Fig.~\ref{fig:preliminaries}a.

For any subset $V' \subset V$, the convex hull of $V'$ is a
\emph{subpolygon} of~$P$.  A subpolygon $Q$ is \emph{proper} if $Q
\neq P$.  We are interested in subpolygons $Q$ of $P$ that closely
approximate $P$. Let $\err(P,Q) \geq 0$ denote the \emph{approximation
error} of $P$ with respect to a subpolygon~$Q$ of~$P$. We will
consider two different error functions: the aperture-angle and the
Hausdorff-distance (to be defined below). Let $\Ck(P)$ denote the set
of convex subpolygons of $P$ with at most $k$ vertices and let
$\errk(P)$ denote the smallest error that can be achieved by a polygon
in $\Ck(P)$, that is,
\[
\errk(P):= \min_{Q\in \Ck(P)} \err(P,Q).
\]
Clearly $\errk(P)= 0$ if $P$ has at most $k$ vertices. We require
that $\err(P,Q)$ can be expressed as
\[
\err(P, Q) = \max_{v \in V} \err(v, Q),
\]
where $\err(v, Q)$ is zero if $v$ is a vertex of $Q$, and is otherwise
of the form
\[
\err(v, Q) = \vst(v, s, t),
\]
where $s$ and $t$ are the two consecutive vertices of $Q$ with $s
\prec v \prec t$.  Furthermore, we require the function $\vst$ to be
\emph{monotone}, that is, if $s \preceq s' \prec v \prec t' \preceq
t$, then $\vst(v, s', t') \leq \vst(v, s, t)$.

If we set $\vst(v,s,t)$ to be the distance of $v$ from the
segment~$st$, then $\err(P,Q)$ is the Hausdorff-distance between $P$
and~$Q$, as shown in Fig.~\ref{fig:preliminaries}b.

If we set $\vst(v,s,t) = \pi - \angle svt$, then $\pi - \err(P,Q)$ is
the smallest aperture angle of any vertex of $P$ with respect to~$Q$,
see Fig.~\ref{fig:preliminaries}c.  Since it is easy to see that
$\aper(x, Q)$ is minimized within~$P$ at the vertices of~$P$, this
implies $\err(P, Q) = \pi - \aper(P, Q)$.  Note that we use the
complement of the angle since we want to minimize the error, but
maximize the angle.


\section{Properties of worst-approximable polygons}
\label{sec:properties}

Let us call a convex polygon $P$ \emph{worst-approximable} if for
every proper subpolygon $Q$ of $P$ we have $\errk(Q) < \errk(P)$.
In this section we study the nature of worst-approximable
polygons.  Our arguments are purely combinatorial, using only the
monotonicity of~$\vst$.

We start by introducing some basic concepts. A pair $(p,q) \in V^{2}$
is called a \emph{diagonal}.  For a given approximation error $\ERR
> 0$, a diagonal $(p,q)$ is called \emph{feasible} if for every $v \in
V$ with $p \prec v \prec q$ we have $\vst(v,p,q) \leq \ERR$.  By
monotonicity of $\vst$, if $(p,q)$ is feasible, then so is any
$(p',q')$ with $p \preceq p' \prec q' \preceq q$. A feasible diagonal
$(p,q)$ is called a \emph{chord} if it is the longest feasible
diagonal starting at~$p$.  The \emph{chord graph}~$\G$ is a directed
graph with vertex set $V$, such that $(p, q)$ is an edge of $\G$ if
$(p,q)$ is a chord.  Clearly, every vertex of $\G$ has out-degree one.

\begin{figure}[htb]
  \centering
  \includegraphics[height=3cm]{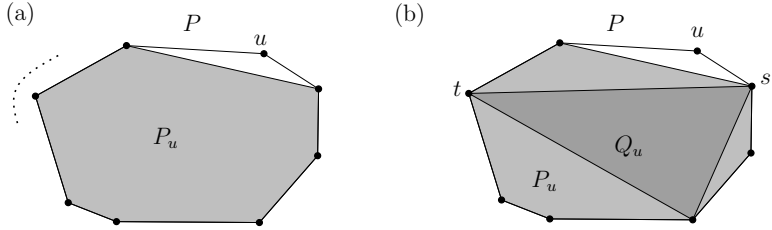}
  \caption{(a) Illustrating $P$ and $P_u$, and (b) $Q_u$ with base $st$.}
  \label{fig:properties}
\end{figure}

Let $k > 2$ and $\ERR > 0$ be fixed, and assume that $P$ is a convex
$n$-gon with $\errk(P) > \ERR$, but such that for every proper
subpolygon $R$ of $P$ we have $\errk(R) \leq \ERR$.  As before, let
$V$ be the set of $P$'s vertices, and let $\G$ be the chord graph of
$P$ for the approximation error~$\ERR$. For $u \in V$, let $P_{u}$ be
the convex hull of $V \setminus \{u\}$, see
Fig.~\ref{fig:properties}a.  By assumption, $P_{u}$ has a $k$-gon
approximation $Q_{u}$ with error $\leq \ERR$. Without loss of
generality, we can assume that $Q_{u}$ consists of $k-1$ chords and of
an edge $st$, where $s \prec u \prec t$, as illustrated in
Fig.~\ref{fig:properties}b. For all $x \in V \setminus \{u\}$ with $s
\prec x \prec t$ we have $\vst(x,s,t)\leq \ERR$, but $\vst(u,s,t) >
\ERR$.

We call $st$ the \emph{base} of $Q_{u}$, and note that $Q_{u}$ is
completely determined by~$t$, since the other $k-1$ vertices can
be found by following $k-1$ arcs from $t$ in the chord graph. Let
$\vs: V \mapsto V$ and $\vt: V \mapsto V$ be functions mapping
$u$ to the clockwise ($\vs$) and counter-clockwise ($\vt$)
endpoints of the base of~$Q_{u}$.

\begin{lemma}
 The function $\vt$ is a bijection.
\end{lemma}
\begin{proof}
  Assume that there exists two subpolygons $Q_{u}$ and $Q_{v}$, $u\neq
  v$, of $P$ that both have base~$st$. Since $st$ is the base of
  $Q_{u}$ and $s\prec v \prec t$, we have $\vst(u,s,t)>\ERR$ and
  $\vst(v,s,t) \leq \ERR$. This, however, contradicts $\vst(v,s,t)>
  \ERR$ due to $st$ being the base of~$Q_{v}$. Thus, the base of
  each $Q_{u}$ is distinct. Since $Q_{u}$ is completely determined by
  $\vt(u)$, the function $\vt$ is an injection, and therefore a
  bijection.
\end{proof}

For a base $st$, let the \emph{witness} $\w(st)$ of $st$ be the unique
vertex with $s \prec \w(st) \prec t$ with $\vst(\w(st),s,t)
>\ERR$. The witness of the base of $Q_{u}$ is of course~$u$.  There
are thus $n$ bases in total, and their witnesses are all distinct.

\begin{lemma} \label{lem:not-nested}
  It is impossible for two bases $st$ and $s't'$ to be nested, that
  is, to realize the order
  \[
  s \preceq s' \prec t' \preceq t.
  \]
\end{lemma}
\begin{proof}
  Assume the opposite, that is, $s \preceq s' \prec t' \preceq t$.
  Since $\w(s't') \neq \w(st)$ we have $\vst(\w(s't'),s,t) \leq
  \ERR$. By monotonicity of $\vst$ this implies $\vst(\w(s't'),
  s',t') \leq \ERR$, a contradiction.
\end{proof}

\begin{figure}[htb]
  \centering
  \includegraphics[width=14cm]{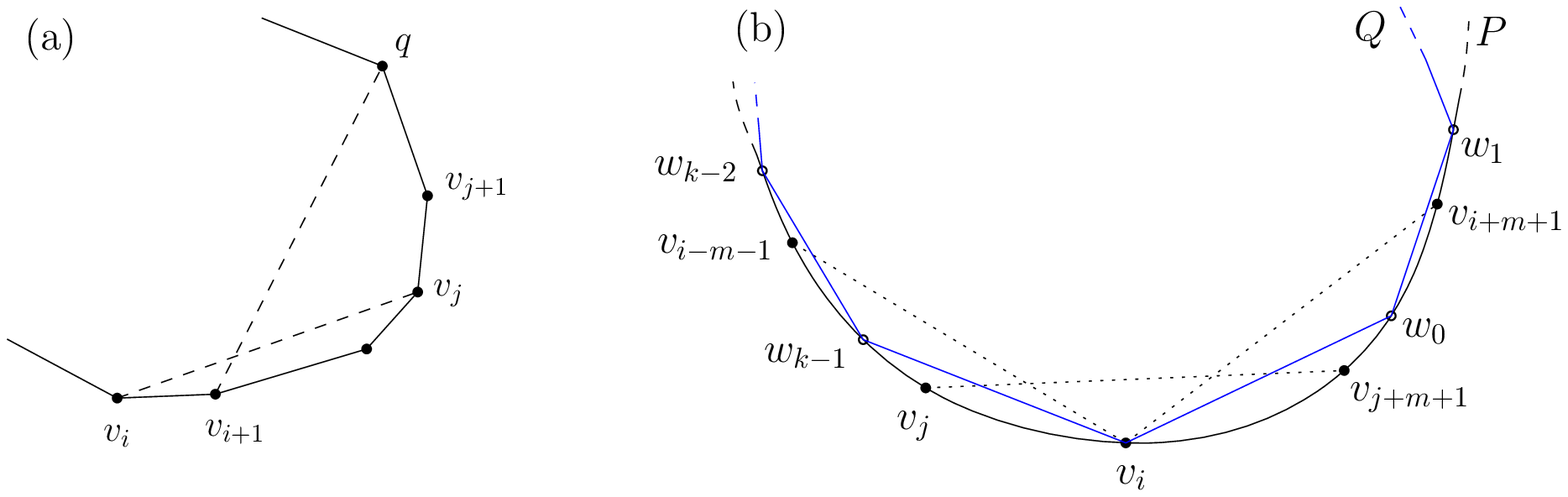}
  \caption{(a) Illustrating the proof of Lemma~\ref{lem:indegree-one},
  and (b) the proof of Lemma~\ref{lem:near-witness}} 
  \label{fig:near-witness}
\end{figure}

Next we study the nature of the chord graph and prove that it is
surprisingly symmetric.  We denote the vertices $V$ of $P$ as $v_{0},
v_{1}, \dots, v_{n-1}$ in counter-clockwise order.  Throughout the
paper, arithmetic on indices of $v$ will be modulo~$n$.
\begin{lemma}
  \label{lem:indegree-one}
  Every vertex in the chord graph has in-degree and out-degree one.
  There is an integer $m > 1$ such that every chord is of the form
  $(v_{i}, v_{i+m})$.
\end{lemma}
\begin{proof}
  From the definition of chord graphs it immediately follows that the
  out-degree of each vertex is one. Since the number of arcs in the
  chord graph is $n$, it suffices to show that no vertex has in-degree
  greater than one.  Assume that $q$ is a vertex with in-degree at
  least two, that is, there are chords $(p,q)$ and $(p', q)$, with $p
  \prec p' \prec q$. Let $u = \vt^{-1}(p)$ and $u' = \vt^{-1}(p')$,
  and so $\vt(u) = p$ and $\vt(u') = p'$.  Since we can find $\vs(u)$
  from $\vt(u)$ by following $k-1$ arcs in the chord graph, $\vt(u) =
  \vt(u')$ implies $\vs(u) = \vs(u')$.  Now we have $s(u')\preceq s(u)
  \preceq t(u) \prec t(u')$, a contradiction to
  Lemma~\ref{lem:not-nested}.  It follows that no vertex of the chord
  graph has in-degree greater than one.

  Next we consider two chords $(v_{i}, v_{j})$ and $(v_{i+1}, q)$, as
  shown in Fig.~\ref{fig:near-witness}a.
  If $q \neq v_{j+1}$ then there has to be a chord $(p, v_{j+1})$
  with $v_{i} \prec p \prec v_{i+1} \prec v_{j} \prec v_{j+1} \prec
  q$, a contradiction  .  This implies that two consecutive chords must
  have the same length, and so all chords do.
\end{proof}

We will use $m$ to denote the chord ``length'' as in the lemma.  For
every $0 \leq i < n$, $(v_{i}, v_{i+m})$ is the chord starting at
$v_{i}$, and $(v_{i-m}, v_{i})$ is the chord ending at~$v_{i}$.

Recall that every $Q_u$ has $k-1$ chords and a base. Since the length of
every chord is~$m$ by Lemma~\ref{lem:indegree-one}, it follows
immediately that every base must have length $n - (k-1)m$:
\begin{corollary} \label{cor:baselength}
  Every base is of the form $(v_{i}, v_{i+n-(k-1)m})$.
\end{corollary}

It turns out that we can prove even stronger properties about the
bases.

\begin{lemma}
  \label{lem:base-length}
  All bases have length $m+1$, and so $n = km+1$.
\end{lemma}
\begin{proof}
  By Corollary~\ref{cor:baselength} all bases have length $b := n -
  (k-1)m$. Assume that $b\neq m+1$. Since a base must be longer than
  a chord it suffices to consider the case when $b > m + 1$.
  Consider two consecutive bases $(v_{0}, v_{b})$ and $(v_{1},
  v_{b+1})$. Consider the diagonal $(v_{1}, v_{b})$.  Since
  $b > m + 1$, this diagonal is not feasible, and so there is a
  vertex $w \in V$ with $v_{1}\prec w \prec v_{b}$ with
  $\vst(w,v_{1},v_{b}) > \ERR$.

  By monotonicity, this implies $\vst(w,v_{0},v_{b}) > \ERR$, and
  $\vst(w,v_{1},v_{b+1}) > \ERR$.  However, $\w(v_{0}v_{b})$ is the
  only vertex in the range $v_{0} \prec x \prec v_{b}$ with
  $\vst(x,v_{0},v_{b}) > \ERR$, and so $w = \w(v_{0}v_{b})$.
  Similarly, $\w(v_{1}v_{b+1})$ is the only vertex in the range
  $v_{1}\prec x \prec v_{b+1}$ with $\vst(x, v_{1}, v_{b+1}) >
  \ERR$, and so $w = \w(v_{1}v_{b+1})$.  It follows that
  $\w(v_{0}v_{b}) = \w(v_{1}v_{b+1})$, a contradiction since all
  witnesses are distinct.
\end{proof}

\begin{lemma}
  \label{lem:near-witness}
  Let $v_{i} \in V$, and consider the two witnesses
  $\w(v_{i-m-1}v_{i})$ and $\w(v_{i}v_{i+m+1})$. If $n > k + 1$ then
  at least one of the two witnesses lies in the range $\vs(v_{i})
  \prec x \prec \vt(v_{i})$.
\end{lemma}
\begin{proof}
  Set $v_{j} := \vs(v_{i})$ and note that $v_{i-m-1} \prec v_{j} \prec
  v_{i} \prec v_{j+m+1} \prec v_{i + m + 1}$, as shown in
  Fig.~\ref{fig:near-witness}b.  Consider the $k$
  witnesses $w_{a} := \w(v_{i+am}v_{i+am+m+1})$, where $0\leq a \leq
  k-1$.  Note that $w_{0} = \w(v_{i}v_{i+m+1})$, and $w_{k-1} =
  \w(v_{i-m-1}v_{i})$ since $v_{i-m-1} = v_{i+(k-1)m}$.

  The proof is done by contradiction, thus assume $v_{i-m-1}
  \prec w_{k-1} \preceq v_{j}$ and $v_{j+m+1}\preceq w_{0}
  \prec v_{i+m+1}$. Now consider the subpolygon $R$ with the
  $k+1$ vertices $\{v_{i}, w_{0}, w_{1},\dots, w_{k-1}\}$. We
  will show that $\errk(R) > \ERR$, which is a contradiction
  to the definition of~$P$.

  First, we observe that $v_{i+am} \prec w_{a} \preceq v_{i+(a+1)m}$
  and that $w_{k-1} \prec v_{i} \prec w_{0}$. This implies that the
  vertices of $R$ are $ v_{i}, w_{0}, w_{1},\dots, w_{k-1}$ in this
  order.  It remains to verify the approximation error.

  From the definition of a witness we have $\vst(w_{a}, v_{i+am},
  v_{i+am+m+1}) > \ERR$ and $w_{a-1} \preceq v_{i+am} \prec
  w_{a} \prec v_{i+am+m+1} \preceq w_{a+1}$,  for $1 \leq a \leq k-2$.
  Putting these two observations together implies that
  $\vst(w_{a}, w_{a-1}, w_{a+1}) > \ERR$ since $\vst$ is a
  monotone function. 

\iffalse
  Using similar arguments it is easy to
  show that $\vst(w_{0}, v_{i}, w_{1})$,
  $\vst(w_{k-1},w_{k-2},v_{i})$, $\vst(v_{i}, v_{j}, v_{j+m+1})$
  and $\vst(v_{i}, w_{k-1},w_{0})$ must all be greater than
  $\ERR$.
\else
  Similarly, since $\vst(w_{0}, v_{i}, v_{i+m+1})
  > \ERR$ and $w_{0} \prec v_{i+m+1} \preceq w_{1}$ we have
  $\vst(w_{0}, v_{i}, w_{1}) > \ERR$.

  Since $\vst(w_{k-1}, v_{i-m-1}, v_{i}) > \ERR$, and $w_{k-2}
\preceq
  v_{i-m-1} \prec w_{k-1}$, we also have $\vst(w_{k-1},w_{k-2},v_{i}) >
  \ERR$.

  Finally, since $v_{i} = \w(v_{j}v_{j+m+1})$, we have $\vst(v_{i},
  v_{j}, v_{j+m+1}) > \ERR$.  Since $w_{k-1} \preceq v_{j} \prec
  v_{i} \prec v_{j+m+1} \preceq w_{0}$, we also have $\vst(v_{i},
  w_{k-1},w_{0}) > \ERR$.
\fi

  Hence, for any subpolygon $Q$ of $R$ with $k$ vertices, the vertex
  $v$ of $R$ not in $Q$ has approximation error $\err(v, Q) > \ERR$,
  implying $\err(R, Q) > \ERR$.
\end{proof}


\section{Aperture-angle approximation}
\label{sec:aperture}

In order to proceed with our arguments, we need to bring back in the
geometry of the problem.  In this section we consider the case of
aperture angle approximation, that is, our error function is
$\vst(v,s,t) = \pi - \angle svt$ as illustrated in
Fig.~\ref{fig:preliminaries}c. 

\begin{figure}[htb]
  \centering
  \includegraphics{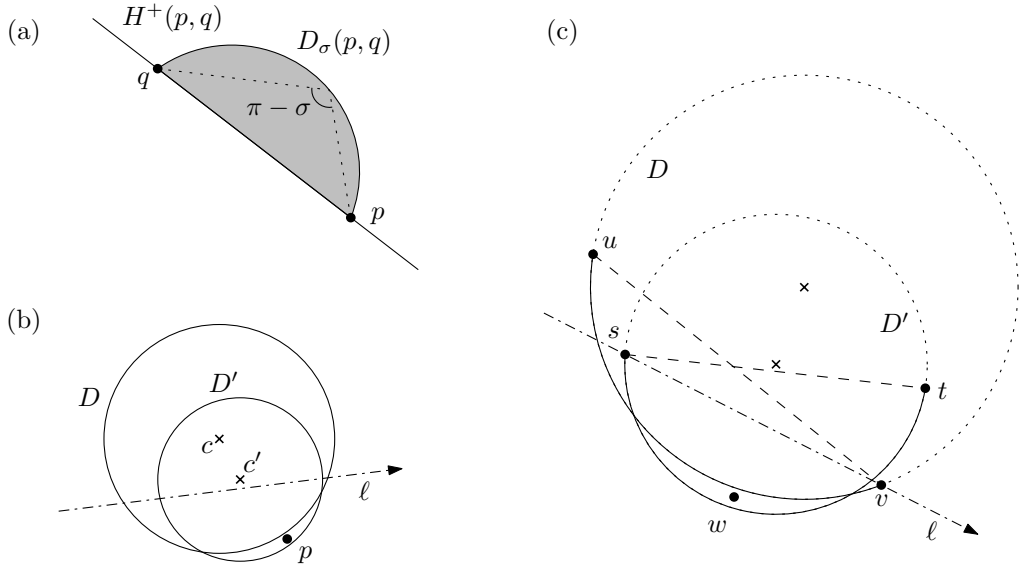}
  \caption{(a) The region $\D(p,q)$. (b) The radius of $D'$ is smaller
    than that of $D$.  (c) $D$ and $D'$ fulfill the assumptions of
    Lemma~\ref{lem:stick-out}.}
  \label{fig:aperture}
\end{figure}

For two points $p$, $q$ in the plane, let $H^{+}(p, q)$ be the
half-plane to the right of the oriented line from $p$ to~$q$. For any
$0 < \ERR < \pi$ we define
\[
\D(p,q) := \{ x \in H^{+}(p,q) \mid \angle pxq \geq \pi - 
\ERR \}.
\]
The region $D_{\ERR}(p, q)$ is the intersection of a disk containing
$p$ and $q$ on its boundary with $H^{+}(p,q)$, see
Fig.~\ref{fig:aperture}a.  Its boundary consists of a circular arc
with endpoints~$p$ and~$q$ and of the segment~$pq$. For fixed~$\ERR$,
the regions $D_{\ERR}(p,q)$ for any pair $p, q$ are affinely
similar---that is, one can be obtained from the other by a rigid
motion and a scaling---and in particular the radius of the circular
arc is proportional to the distance~$d(p, q)$.  If $0 < \ERR < \pi/2$,
then the center of the circular arc lies to the left of the oriented
line~$pq$, and so the circular arc spans less than a semi-circle.  We
observe that $\vst(v, s, t) \leq \ERR$ if and only if $v \in \D(s,t)$
(note that $s \preceq v \preceq t$ implies that $v \in H^{+}(s, t)$).

We will need a simple geometric fact, which can be proven using
elementary calculations, see Fig.~\ref{fig:aperture}b.
\begin{lemma}
  \label{lem:stick-out}
  Consider two disks $D$ and $D'$, whose centers lie to the left of an
  oriented line~$\ell$.  If $D' \cap \ell$ is contained in $D\cap
  \ell$, but there is a point $p \in D' \setminus D$ to the right of
  $\ell$, then the radius of $D'$ is smaller than the radius of~$D$.
\end{lemma}

\begin{lemma}
  \label{lem:min-aa}
  If $P$ is a convex $n$-gon that is worst-approximable with respect
  to the aperture angle and $k>2$ then $n = k+1$.
\end{lemma}
\begin{proof}
  Let $P$ be a worst-approximable convex $n$-gon with respect to the
  aperture angle, and assume that the statement of the lemma is false,
  that is, $n>k+1$.  Let $\ERR := \max_{R} \errk(R)$, where the
  maximum is taken over all proper subpolygons~$R$ of~$P$. Since $P$
  is worst-approximable, we have $\errk(P) > \ERR$, and since for
  every proper subpolygon $R$ of $P$ we have $\errk(R) \leq \ERR$, the
  results of Section~\ref{sec:properties} apply.  Since $\aper(3) =
  \pi/2$~\cite{bmp-rpdg-05}, we have $\ERR < \errk(P) \leq \pi/2$.

  Let now $v$ be a vertex of $V$ that maximizes the Euclidean distance
  $d(\vs(v), \vt(v))$.  According to Lemma~\ref{lem:near-witness}
  there is an incident base, say $uv$, such that $u \prec \vs(v) \prec \w(uv)
  \prec v$ (the other case being symmetric).  We let $w := \w(uv)$, $s
  := \vs(v)$, $t := \vt(v)$, and consider the sequence of five
  vertices $u \prec s \prec w \prec v \prec t$.

  Let $D$ be the disk supporting $\D(u,v)$, and let~$D'$ be the disk
  supporting~$\D(s, t)$, see Fig.~\ref{fig:aperture}c.  By our
  observation above, the only vertex in the range $u\preceq x \preceq
  v$ not in $D$ is $w$, and the only vertex in the range $s \preceq x
  \preceq t$ not in $D'$ is~$v$.  In particular, we have $s \in D$,
  $w\not\in D$, $w\in D'$, and $v \not\in D'$, see
  Fig.~\ref{fig:aperture}c.

  Let $\ell$ be the directed line from $s$ to~$v$.  Since $\ERR <
  \pi/2$, the center of $D$ lies to the left of the oriented line
  $uv$, and therefore to the left of~$\ell$.  Similarly, the center of
  $D'$ lies to the left of $st$, and therefore to the left of~$\ell$.
  Since $s \in D$ and $v \not\in D'$, we have $D' \cap \ell \subset
  D\cap \ell$.  Furthermore, since $s \prec w \prec v$, the point $w$
  lies to the right of~$\ell$, and in $D' \setminus D$.

  By Lemma~\ref{lem:stick-out} the radius of $D'$ is therefore less
  than the radius of~$D$.  Since $\D(u,v)$ and $\D(s, t)$ are affinely
  similar, this implies that $d(s, t) < d(u, v)$.  This, however, is a
  contradiction to our choice of~$v$, and our assumption $n > k+1$ is
  false.
\end{proof}

A limit argument now gives the following theorem:
\begin{theorem}
  \label{thm:main}
  For any compact convex set $C \subset \R^{2}$ and any integer $k >
  2$ there is a convex $k$-gon $Q$ contained in $C$ such that
  $\aper(C, Q) \geq \aper_k$, where $\aper_{k} =
  \big(1-2/(k+1)\big)\pi$.  This bound is best possible.
\end{theorem}
\begin{proof}
  We start by proving the theorem for the special case when $C$ is a
  convex polygon.  Among all subpolygons $R$ of $C$ with $\errk(R)
  \geq \errk(C)$, let $P$ be one with the minimal number of vertices.
  This implies that for every proper subpolygon $R$ of $P$ we have
  $\errk(R) < \errk(C) \leq \errk(P)$, and so $P$ is
  worst-approximable.  By Lemma~\ref{lem:min-aa}, $P$ is a
  $(k+1)$-gon.  It follows that $P$ has at least one interior angle
  that is at least $\aper_{k}$.  Choosing $Q$ to be the convex hull of
  the remaining $k$ vertices gives $\err(P, Q) \leq \pi - \aper_{k}$, and so
  $\errk(C) \leq \errk(P) \leq \pi - \aper_{k}$, proving the theorem.

  Next we consider a general compact convex figure $C$ in the plane.
  We choose a sequence $P_{i}$ of convex polygons inscribed within
  $C$ that converges to $C$ with respect to the Hausdorff-distance.
  For each $P_{i}$ there is a subpolygon $Q_{i} \subset P_{i}$ with
  $k$ vertices and $\aper(P_{i}, Q_{i}) \geq \aper_{k}$.

  We interpret the $k$ vertices of $Q_{i}$ as a point $q_{i} \in
  \R^{2k}$.  Since $Q_{i} \subset C$, this sequence is bounded, and so
  the Bolzano-Weierstrass Theorem guarantees the existence of a
  subsequence that converges to a point $q \in \R^{2k}$.  We interpret
  $q$ again as a $k$-vertex polygon~$Q$.  It is easy to see that $Q$
  is a convex polygon with at most $k$ vertices.

  It remains to show that $\aper(C, Q) \geq \aper_{k}$.  Let $p\in C
  \setminus Q$. There is a sequence of points $p_{i} \in P_{i}$ with
  $\lim_{i\rightarrow \infty} p_{i} = p$.  Since $\aper(P_{i},
  Q_{i})\geq \aper_{k}$, that implies that there are vertices $x_{i},
  y_{i}$ of $Q_{i}$ such that $\angle x_{i}p_{i}y_{i} \geq \aper_{k}$.
  We consider the sequence $(x_{i}, y_{i})$ in $\R^{4}$ and apply
  again the Bolzano-Weierstrass Theorem.  We pass to a subsequence
  where $x := \lim_{i\rightarrow \infty} x_{i}$ and $y :=
  \lim_{i\rightarrow \infty} y_{i}$ exist.  The points $x$ and $y$ are
  necessarily vertices of $Q$.  The angle $\angle xpy$ is a continuous
  function in $(x, y, p)$ as long as the three points remain distinct.
  Since $p \not\in Q$, this implies $\angle xpy \geq \aper_{k}$, and
  the theorem follows.

  The regular $(k+1)$-gon shows that the bound is indeed best
  possible.
\end{proof}

Theorem~\ref{thm:main} implies that $\aper(k) = \aper_{k}$, positively
answering Conjecture~\ref{con:aperture} (the case $k=2$ was already
known to be true).


\section{Hausdorff-approximation}
\label{sec:hausdorff}

In this section we consider the case of Hausdorff approximation, and
our error function $\vst$ is the distance between $v$ and the
segment~$st$, as illustrated in Fig.~\ref{fig:preliminaries}b.  We
continue with the analysis of worst-approximable polygons of
Section~\ref{sec:properties}: $k > 2$ and $\ERR > 0$ are fixed, and we
consider a convex $n$-gon $P$ with $\errk(P) > \ERR$, but such that
for every proper subpolygon $R$ of $P$ we have $\errk(R) \leq \ERR$.

We need a small geometric result similar to
Lemma~\ref{lem:not-nested}: 
\begin{lemma}
  \label{lem:base-order}
  With respect to the Hausdorff-approximation, it is impossible for
  two bases $st$ and $s't'$ to realize the order $s \prec s' \prec
  \w(s't') \prec \w(st) \prec t \prec t'$.
\end{lemma}
\begin{proof}
  Suppose that there exist two bases $st$ and $s't'$ such that $s
  \prec s' \prec \w(s't') \prec \w(st) \prec t \prec t'$.  Let $w :=
  \w(st)$, let $w' := \w(s't')$, let $q$ be the point on the segment
  $st$ minimizing the distance~$d(w', q)$, and let $q'$ be the point
  on $s't'$ minimizing~$d(w, q')$.
  
  \begin{figure}[htb]
    \centering
    \includegraphics{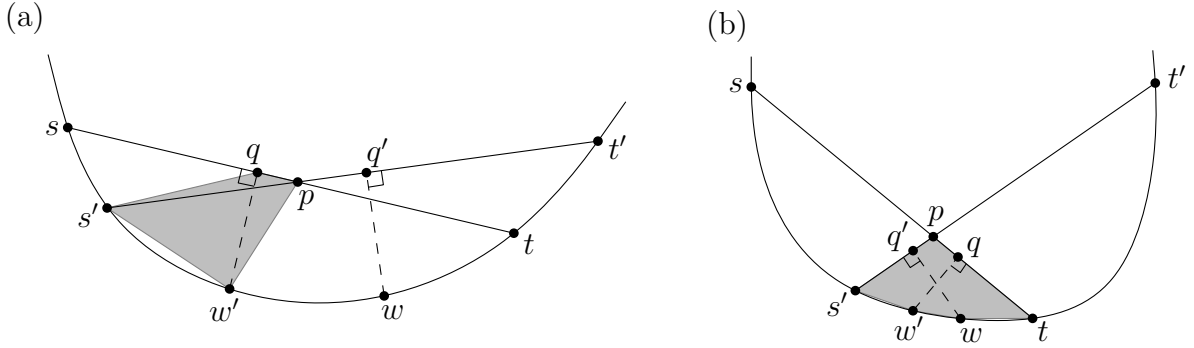}
    \caption{Proof of Lemma~\ref{lem:base-order}.}
    \label{fig:base-order}
  \end{figure}
  The bases $st$ and $s't'$ must intersect in a point~$p$ lying
  in~$P$. We first assume that $q$ lies on the segment~$sp$, see
  Fig.~\ref{fig:base-order}a.  Since $ss'w't$ is a convex
  quadrilateral, so is $pqs'w'$.  Its diagonals $s'p$ and $w'q$
  intersect, implying $d(w', s't') < d(w',q) \leq \ERR$, a
  contradiction to $w' = \w(s't')$.  It follows that $q$ must lie on
  the segment $pt$, and by symmetry $q'$ lies on $s'p$.

  Since $ps'w'wt$ is a convex pentagon (the intersection of the two
  subpolygons $ss'w'wt$ and $s'w'wtt'$ of~$P$), $q$ lies on its edge
  $pt$, and $q'$ lies on its edge $s'p$, the chain $q'w'wq$ is a
  convex quadrilateral, see Fig.~\ref{fig:base-order}b.  The sum of
  the lengths of its diagonals $q'w$ and $qw'$ must be larger than the
  sum of the lengths of the opposite sides $q'w'$ and $qw$. This,
  however, is a contradiction to $d(w, q') \leq \ERR < d(w, q)$ and
  $d(w', q) \leq \ERR < d(w', q')$, and the lemma follows.
\end{proof}

\begin{lemma}
  \label{lem:r}
  There is an integer $0 < r \leq m$ such that for every $v_{i}$ we
  have $\vs(v_{i}) = v_{i-r}$ .
\end{lemma}
\begin{proof}
  Let $r$ be the smallest integer $> 0$ such that there is a vertex
  $v_{i}$ with $\vs(v_{i}) = v_{i-r}$.  We will show that then
  $\vs(v_{i+1})=v_{i-r+1}$, and by induction this implies the lemma.
  Assume the opposite, that is, $\vs(v_{i+1}) \neq v_{i-r+1}$.  By
  definition of $r$, we cannot have $v_{i-r+1} \prec \vs(v_{i+1})
  \prec v_{i+1}$, and therefore $\vs(v_{i+1}) \prec \vs(v_{i}) \prec
  v_{i} \prec v_{i+1} \prec \vt(v_{i+1}) \prec \vt(v_{i})$, which is a
  contradiction to Lemma~\ref{lem:base-order}.
\end{proof}

Note that the above lemma also implies $\vt(v_{i}) = v_{i-r+m+1}$.
From now on, let $r$ be as in Lemma~\ref{lem:r}.  We may assume $r
\leq m/2$, otherwise we can work with the mirror image of~$P$.

The rest of the proof is similar in spirit to Lemma~\ref{lem:min-aa},
but is technically more complicated.  
\begin{lemma}
  \label{lem:3r}
  We have $3r > m+1$.
\end{lemma}
\begin{proof}
  Assume $3r \leq m+1$, and let $v_{i}$ be a vertex minimizing the
  Euclidean distance $d(v_{i}, v_{i-r})$.  We concentrate on the
  vertices
  \[
  s_{1} = v_{i-2r}, \quad
  s_{2} = v_{i-r}, \quad
  s_{3} = v_{i}, \quad
  s_{4} = v_{i+r}, \quad
  t_{1} = v_{i-2r+m+1}, \quad
  t_{2} = v_{i-r+m+1}, \quad
  t_{3} = v_{i+m+1}.
  \]
  Since $3r \leq m+1$ we have $s_{1} \prec s_{2} \prec s_{3} \prec
  s_{4} \preceq t_{1} \prec t_{2}$.  We have $s_{2} = \w(s_{1}t_{1})$,
  $s_{3} = \w(s_{2}t_{2})$, and $s_{4} = \w(s_{3}t_{3})$. By our
  choice of $v_{i}$, we have $d(s_{2}, s_{3}) \leq d(s_{1}, s_{2})$
  and $d(s_{2}, s_{3}) \leq d(s_{3}, s_{4})$.  We will show that this
  is impossible, implying that the assumption $3r \leq m+1$ is false.

  \begin{figure}[htb]
    \centering
    \includegraphics[width=\textwidth]{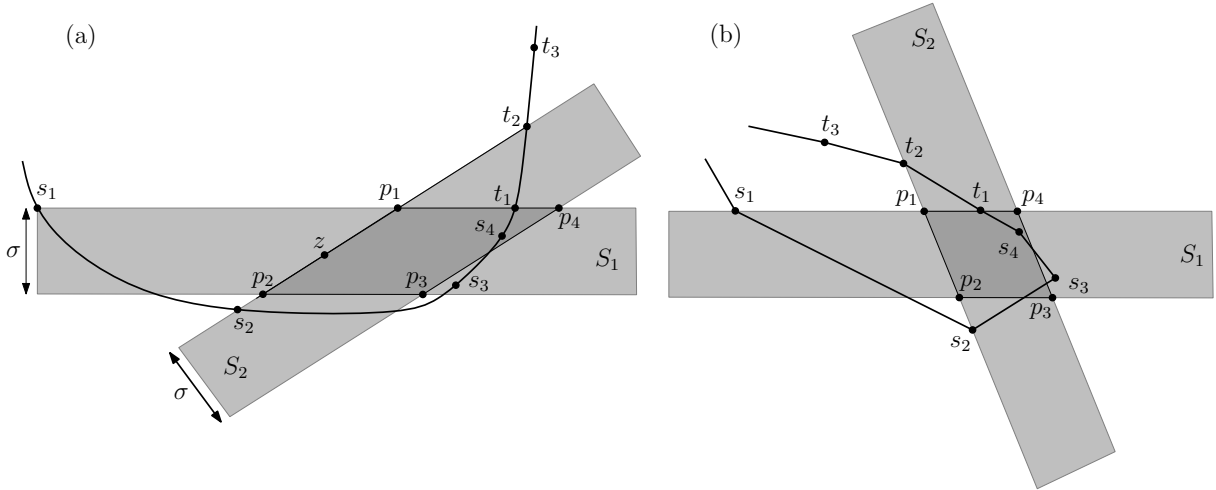}
    \caption{Proof of Lemma~\ref{lem:3r}.}
    \label{fig:3r}
  \end{figure}
  By rotating $P$, we can assume that the line $s_{1}t_{1}$ is
  horizontal.  Let $S_{1}$ be the horizontal strip of width~$\ERR$
  bounded from above by $s_{1}t_{1}$, and let $S_{2}$ be the strip of
  width~$\ERR$ to the right of the oriented line from $s_{2}$ to
  $t_{2}$.  The intersection $R = S_{1} \cap S_{2}$ is a rhombus.  Its
  top-left corner $p_{1}$ is the intersection point of the segments
  $s_{1}t_{1}$ and $s_{2}t_{2}$.  We denote the remaining corners
  counter-clockwise as $p_{2}$, $p_{3}$, and~$p_{4}$, see
  Fig.~\ref{fig:3r}.  Since $s_{1} \prec s_{3} \prec s_{4} \preceq
  t_{1}$, we have $s_{3}, s_{4}, t_{1} \in S_{1}$.  Since we also have
  $s_{4}, t_{1} \in S_{2}$, it follows that $s_{4}, t_{1} \in R$.

  We now first show that $s_{4}$ must lie strictly right of $s_{3}$.
  If $t_{1}$ lies right of $s_{3}$, this follows from the fact that
  $s_{4}$ lies strictly to the right of the oriented line $s_{3}t_{1}$
  and to the left of the oriented line $s_{1}s_{3}$.  If $t_{1}$ lies
  left of $s_{3}$, then $d(s_{3},s_{1}t_{1}) \leq \ERR$ implies
  $d(s_{3}, t_{1}) \leq \ERR$.  If $s_{4}$ lies not strictly right of
  $s_{3}$, then the angle $\angle s_{3}s_{4}t_{1}$ is right or obtuse,
  and so $d(s_{3},s_{4}) \leq d(s_{3}, t_{1}) \leq \ERR$, a
  contradiction to $s_{4} = \w(s_{3}t_{3})$.
  
  We will now first consider the case that the angle $\angle
  s_{2}p_{1}t_{1}$ is at least $90^{\circ}$, so $R$ has obtuse angles
  at $p_{1}$ and $p_{3}$ and acute angles at $p_{2}$ and $p_{4}$, see
  Fig.~\ref{fig:3r}a.  Let $z$ be the point on the segment
  $p_{1}p_{2}$ at distance $\ERR$ from~$p_{1}$.  Since $s_{2} =
  \w(s_{1}t_{1})$, $s_{2}$ must lie below~$z$, and so the segment
  $zp_{1}$ is a subset of $s_{2}t_{2}$.  Since $s_{3}$ lies in $S_{1}$
  but $d(s_{3},s_{2}t_{2}) > \ERR$, $s_{3}$ must lie strictly to the
  right of~$p_{3}$.  Since $s_{4} \in R$ and strictly right of
  $s_{3}$, this implies that $s_{3}$ lies strictly left of $p_{4}$.
  Since $s_{3}$ lies strictly in between the vertical lines through
  $p_{3}$ and $p_{4}$, and $s_{4}$ lies in between the vertical lines
  through $s_{3}$ and $p_{4}$, we have $d(s_{3}, s_{4}) < d(p_{3},
  p_{4})$.

  Consider now the point~$s_{2}$.  We argue that it must lie strictly
  below~$p_{2}$.  Indeed, otherwise it lies inside $S_{1}$ on the
  segment $p_{1}p_{2}$.  Since $d(s_{2}, s_{1}t_{1}) > \ERR$, this
  implies that $s_{1}$ lies right of $s_{2}$. Then the angle $\angle
  s_{2}s_{1}p_{1}$ is obtuse, implying $d(s_{2},s_{1}) \leq
  d(s_{2},p_{1})$.  But since $s_{3}$ lies strictly right of $p_{3}$,
  we have $d(s_{2}, s_{3}) > d(s_{2}, p_{1}) \geq d(s_{2},s_{1})$, a
  contradiction to our choice of~$s_{3} = v_{i}$.

  Since $s_{2}$ lies strictly below $p_{2}$ on the line $p_{1}p_{2}$
  and $s_{3}$ lies in $S_{1}$ right of~$p_{3}$, we have
  $d(s_{2},s_{3}) > d(p_{2},p_{3}) = d(p_{3}, p_{4}) > d(s_{3},
  s_{4})$, again a contradiction to our choice of~$s_{3} = v_{i}$.

  It remains to consider the case that the angle $\angle
  s_{2}p_{1}t_{1}$ is less than $90^{\circ}$, so $R$ has acute angles
  at $p_{1}$ and $p_{3}$ and obtuse angles at $p_{2}$ and $p_{4}$, see
  Fig.~\ref{fig:3r}b.  If $s_{3}$ does not lie in $R$, we immediately
  have a contradiction: Since $s_{4}$ must lie to the left of the
  oriented line $t_{1}s_{3}$ and right of $s_{3}$, it cannot lie
  in~$R$.  If $s_{3}$ lies in $R$, the nearest point to $s_{3}$ on the
  segment $s_{2}t_{2}$ must be $s_{2}$, as otherwise $d(s_{3},
  s_{2}t_{2}) \leq \ERR$.  This implies that $s_{3}$ lies below the
  line $\ell$ through $s_{2}$ orthogonal to $p_1p_2$, and outside the
  circle $T$ with radius~$\ERR$ around~$s_{2}$.
  
  Since the segments $s_{2}s_{4}$ and $s_{3}t_{3}$ intersect and
  $s_{4} = \w(s_{3}t_{3})$, we must have $d(s_{2}, s_{4}) > \ERR$.
  This implies that the nearest point to $s_{4}$ on $s_{2}t_{2}$ must
  be different from $s_{2}$, and so $s_{4}$ must lie above the line
  $\ell$.
  
  Now we observe that if $s_{3}$ lies above $s_{2}$, then it lies
  strictly right of the intersection point of $\ell$ and the
  line $p_4p_3$. Since $s_{4}$ lies right of $s_{3}$, this implies
  that $s_{4}$ cannot lie in $R$ above $\ell$, a contradiction, and so
  $s_{3}$ must lie below~$s_{2}$.  Since $s_{3} \in R$, this implies
  $s_{2}\in R$. Therefore the point nearest to $s_{2}$ on $s_{1}t_{1}$
  must be $t_{1}$, and $t_{1}$ must lie left of~$s_{2}$.  This implies
  that the angle $\angle t_{1}s_{2}s_{3}$ is obtuse, and so
  $d(t_{1},s_{3}) > d(t_{1},s_{2}) > \ERR$.  Since $s_{3}$ lies right
  of $t_{1}$, this implies $d(s_{3}, s_{1}t_{1}) > \ERR$, a contradiction.
\end{proof}

\begin{theorem}
  \label{thm:min-hd}
  If $P$ is a convex $n$-gon that is worst-approximable with respect
  to the Hausdorff-distance and $k>2$ then $n = k+1$.
\end{theorem}
\begin{proof}
  We assume $n>k+1$ and set $\rho := m+1-r$. Since $r \leq m/2$, we
  have $2\rho > m + 1$.  Let $R$ be the subpolygon of $P$ formed by
  the vertices $v_{0}, v_{\rho}, v_{2\rho}, \dots, v_{\ell\rho}$,
  where $\ell$ is an integer such that $(\ell+1)\rho \leq km+1 <
  (\ell+2)\rho$.  Since $\rho > 1$, $R$ is a proper subpolygon of~$P$.
  We will show that $\errk(R) > \ERR$, a contradiction to the
  assumption that $P$ is worst-approximable.

  We first show that $\ell \geq k$.  We assume the contrary, that is
  $\ell < k$. This implies $\ell + 2\leq k+1$. Since $km+1 <
  (\ell+2)\rho$ we have $km+2 \leq (\ell+2)\rho$.  By
  Lemma~\ref{lem:3r} we have $3r > m+1$, which implies $3r \geq m+2$,
  and therefore $\rho \leq \frac 23 m + \frac 13$.  This gives
  \[
  3km + 6 = 3(km+2) \leq 3(\ell+2)\rho \leq 3(k+1)(\frac 23 m + \frac 13)
  = 2km + 2m + k + 1.
  \]
  Rearranging the terms gives $km - 2m -k + 5 \leq 0$ or $(k-2)(n-1)
  \leq -3$, a contradiction with $k > 2$ and $m \geq 1$. It follows
  that our claim holds, that is~$\ell \geq k$.

  It remains to prove that $\errk(R) > \ERR$.  Note that $R$ has at
  least $k+1$ vertices, and so a $k$-vertex subpolygon $Q$ of $R$ must
  exclude at least one vertex of~$R$. We have three cases (recall that
  arithmetic on the indices of $v$ is modulo~$n$):

  \begin{itemize}
  \item $v_{i\rho}$ is excluded, for $0 < i < \ell$\\ By
    Lemma~\ref{lem:r}, $v_{i\rho}$ is the witness of the base
    $(v_{(i+1)\rho-(m+1)}, v_{(i+1)\rho})$.  Since $2\rho > m + 1$,
    $v_{(i-1)\rho} \prec v_{(i+1)\rho - (m+1)} \prec v_{i\rho} \prec
    v_{(i+1)\rho}$, and so monotonicity of $\vst$ implies that
    $\vst(v_{i\rho}, v_{(i-1)\rho}, v_{(i+1)\rho}) > \ERR$.

  \item $v_{0}$ is excluded\\ $v_{0}$ is the witness of the base
    $(v_{\rho-(m+1)}, v_{\rho})$. Since $(\ell+1)\rho \leq km+1 = n$
    and $2\rho > m + 1$ we have $\ell\rho \leq n - \rho < n + \rho -
    (m+1)$, which implies $v_{\ell\rho} \prec v_{\rho-(m+1)} \prec
    v_{0} \prec v_{\rho}$, and so $\vst(v_{0}, v_{\ell\rho}, v_{\rho}) > \ERR$.
    
  \item $v_{\ell\rho}$ is excluded\\ $v_{\ell\rho}$ is the witness of
    the base $(v_{(\ell+1)\rho-(m+1)} v_{(\ell+1)\rho})$.  Since
    $2\rho > m+1$ and $(\ell+1)\rho \leq n$ we have $v_{(\ell-1)\rho}
    \prec v_{(\ell+1)\rho-(m+1)} \preceq v_{0}$, and so
    $\vst(v_{\ell\rho}, v_{(\ell-1)\rho}, v_{0}) > \ERR$.
  \end{itemize}
  In all cases, $\err(R, Q) > \ERR$, and so $\errk(R) > \ERR$.
\end{proof}

The following approximation result is a direct application of
Theorem~\ref{thm:min-hd}.
\begin{theorem}
  \label{thm:approx}
  For any convex polygon $P$ of perimeter at most one and any $k > 2$
  there exists a subpolygon $Q$ of $P$ with $k$ vertices such that
  $\err(P, Q) \leq \frac{1}{k+1}\sin\frac{\pi}{k+1}$ with respect to
  the Hausdorff-distance.  If $P$ is a regular $(k+1)$-gon, this bound
  is best possible.
\end{theorem}
\begin{proof}
  Let $R$ be a subpolygon of $P$ with the smallest number of vertices
  such that $\errk(R) \geq \errk(P)$.  Then $R$ is worst-approximable,
  and by Theorem~\ref{thm:min-hd}, $R$ is a $(k+1)$-gon.  The following
  lemma now implies the theorem.
\end{proof}

\begin{lemma}
  \label{lem:perimeter}
  Let $k > 2$ and let $P$ be a convex $(k+1)$-gon with $\errk(P) = 1$
  with respect to the Hausdorff-distance.  Then the perimeter of $P$
  is at least~$(k+1)/\sin(\pi/(k+1))$, and this bound is tight for the
  regular $(k+1)$-gon.
\end{lemma}
\begin{proof}
  We set $n = k+1$, and let $P$ be a convex $n$-gon with $\errk(P) =
  1$ of minimal perimeter (the existence of such a~$P$ follows from a
  compactness argument).  Let $v_{1},\dots,v_{n}$ denote the vertices
  of $P$ in counter-clockwise order, and let $Q_{i}$ be the subpolygon
  excluding the vertex~$v_{i}$.  We first argue that $\err(P, Q_{i}) =
  1$ for all~$i$.  Indeed, if there is a vertex $v_{i}$ such that
  $\err(P, Q_{i}) > 1$, then we can move $v_{i}$ slightly along the
  directed line from $v_{i}$ to $v_{i+2}$. This decreases the
  perimeter while keeping $\errk(P) = 1$, a contradiction to the
  choice of~$P$.

  Let $\gamma_i$ be the angle made by the oriented diagonals
  $v_{i-1}v_{i+1}$ and $v_i v_{i+2}$.  Since the direction of the
  diagonal $v_{i-1}v_{i+1}$ is a tangent direction at $v_{i}$, we have
  \[
  \sum_{i=1}^{n} \gamma_i = 2\pi.
  \]
  \begin{figure}[htb]
    \centering
    \includegraphics{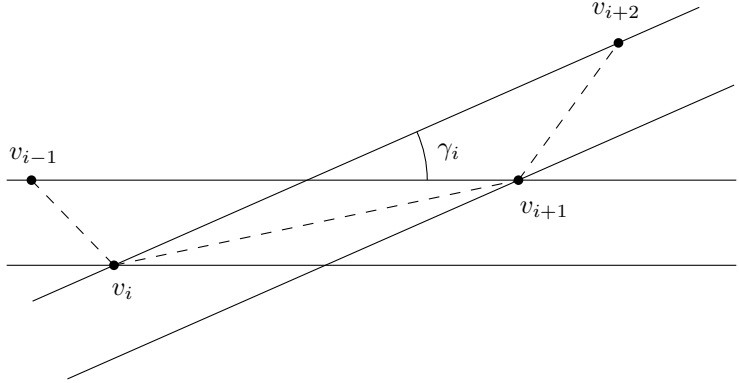}
    \caption{Proof of Lemma~\ref{lem:perimeter}.}
    \label{fig:perimeter}
  \end{figure}
  The distance of $v_{i}$ from the line $v_{i-1}v_{i+1}$ and the
  distance of $v_{i+1}$ from the line $v_{i}v_{i+2}$ is one.  This
  implies that the length of the edge $v_{i}v_{i+1}$ is
  $1/\sin(\gamma_{i}/2)$, see Figure~\ref{fig:perimeter}.  We set $x_i
  = \gamma_i/2$, and define $f(x) = 1/\sin x$.  We then have
  \begin{gather*}
    \sum_{i=1}^{n} x_i  = \pi,\\
    \peri(R) = \sum_{i=1}^{n} f(x_i).
  \end{gather*}
  Since $f''(x) > 0$ for $0 < x < \pi/2$, the function $f(x)$ is
  convex on the interval $(0, \pi/2)$.  Since $0 < x_{i} < \pi/2$, we
  can thus apply Jensen's inequality to give us
  \[ 
  f\Bigl(\frac{\sum_i x_i}{n}\Bigr) \leq \frac{\sum_i f(x_i)}{n} =
  \frac{\peri(R)}{n}
  \]
  Multiplying by $n$ gives $\peri(R) \geq n f(\pi/n) = n/\sin(\pi/n)$,
  completing the proof.
\end{proof}

\section{Conclusions}

We showed that worst-approximable polygons are $(k+1)$-gons for both
the Hausdorff-distance and the aperture-angle distance.  A large part
of the argument is purely combinatorial, using only the monotonicity
of the distance function~$\vst$.  To finish the argument, however, we
needed to make use of some geometric properties of the two distance
functions we considered; in Lemma~\ref{lem:min-aa} for the aperture
angle, and in Lemmas~\ref{lem:base-order} and~\ref{lem:3r} for the
Hausdorff distance.  We must ask: are we just blinded by the geometry
to overlook an entirely combinatorial proof that would apply for any
monotone distance function?

We gave one application of Theorem~\ref{thm:min-hd}, but it's worth
pointing out that the theorem is really far more general, and applies
to many families $\FAM$ of convex polygons.  In many cases the regular
$(k+1)$-gon appears to be the worst case, but this is not always the
case, for instance because the family $\FAM$ does not contain
it. (Consider, for instance, the family of all convex polygons with
vertices on a given ellipse.)

\section{Acknowledgments}

We thank S{\'a}ndor Fekete for the historical background of the
aperture angle problem, and Peter Brass for pointing out the
similarity in the statement of the two conjectures as well as his help
in accessing the literature.  We thank Peter Brass, Hyeon-Suk Na, and
Chan-Su Shin for helpful discussions during a mini-workshop at
Soongsil University.  Finally, we thank Mira Lee for suggesting the
approach used in the proof of Theorem~\ref{thm:approx}.

\bibliographystyle{plain}
\bibliography{angular}
\end{document}